# MECHANICAL DESIGN PRINCIPLES OF AVIAN EGGSHELLS FOR SURVIVABILITY

Fan Liu, Xihang Jiang, and Lifeng Wang*

Department of Mechanical Engineering, Stony Brook University, Stony Brook, NY 11794, USA

* Corresponding authors. E-mail address lifeng.wang@stonybrook.edu

## HIGHLIGHTS

A special 2-crack fracture pattern induced by the dome-shaped structure makes the eggshell harder to break from the outside than from the inside.

The eggshell membrane significantly improves the toughness of the eggshell and makes the toughness tunable.

Maximizing survivability by balancing the protection and nutrient consumption could be a possible driving force in the evolution of eggshell thickness.

## PROGRESS AND POTENTIAL

Biological materials exhibit complex structure-property relationships which are only beginning to be elucidated. Understanding the underlying physical mechanisms of the structure-property relationships is the key to designing bioinspired materials. The eggshell is an excellent example because many design trade-offs are well balanced by its seemingly simple but highly evolved structures. The first design trade-off: to break an eggshell, the internal breaking load needed is much lower than the external breaking load. The reason is that the dome-shaped structure leads to a special stress distribution which inhibits the propagation of the first crack. The second design trade-off: the eggshell is tough during the incubation but weakened at the time of hatching. The membrane of the eggshell significantly improves the toughness by bridging the primary crack and creating more secondary cracks. Moreover, the moisture content-controlled property of the membrane makes the toughness of the eggshell tunable. The third design trade-off: the eggshell must be thick enough to withstand the external forces, but the supply of the building material calcium is limited. The proposed three-index model indicates that the driving force in the evolution of eggshell thickness could be the balance between the protection and nutrient consumption to maximize survivability. The general methodologies presented here hold great potential for the development of eggshell-inspired structural materials that can be used in sports safety applications and the packaging industry.

**MANUSCRIPT CATEGORIZATION (MAP)**

Understanding

**GRAPHICAL ABSTRACT**

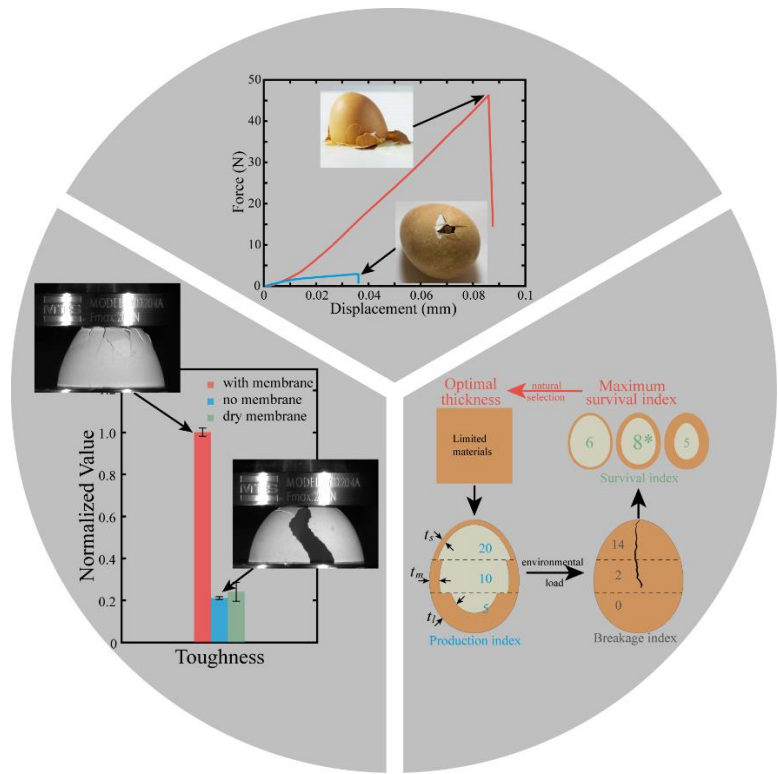

**SUMMARY**

Calcified eggshell was one of the key evolutionary innovations that enabled amniotes to flourish in the dry land. Eggshell has multiple functions to support the incubation, and therefore it has several design trade-offs that must be balanced. First, it must be sufficiently strong to resist the load applied from the outside, meanwhile, it must be easy to break from the inside. Second, the eggshell must be tough to delay the possible crack growth during the incubation, but the toughness must be weakened at the time of hatching. Third, the eggshell must be thin to save the calcium that is limited in the natural diet, but not too thin to protect the embryo from the physical environment. All these trade-offs are balanced by the delicate design of the eggshell, which is the result of millions of years of evolution. In this paper, we uncover the three underlying physical mechanisms of the eggshell structure to balance the three trade-offs by performing experimental tests and simulations. We discovered two distinct fracture patterns that account for the difference in breaking forces between outside load and inside load. We studied the effect of the membrane on the toughness of the eggshell and proved that the decrease of moisture content of the membrane in the incubation weakened the eggshell. We proposed a three-index model which revealed the relationship between the eggshell thickness and offspring number. The three-index model indicates that the driving force in the evolution of eggshell thickness could be the balance between the protection and nutrient consumption to maximize survivability.

**INTRODUCTION**

Egg-laying reproduction was one of the most successful evolutionary innovations that enabled vertebrate animals to leave the water and conquer dry land ~360 million years ago[1]. Calcified eggshell is produced by all modern birds and it is thought to have played an important part in the survival of birds through the Cretaceous–Palaeogene extinction[2]. The calcified eggshell is an important structure to the developing embryo for three reasons. It protects the embryo from the microbial and physical environment. Additionally, it regulates the water and gas exchange during the incubation. The calcified eggshell also serves as a calcium store and provides calcium for the embryo's skeleton growth[3]. The diverse functions of the eggshell result from its hybrid organic-inorganic composite materials and the sophisticated structures that developed in million years of evolution.

Avian eggshell is a complex bioceramic comprising a calcium-carbonate mineral constituent (~95% by weight) and an organic matrix (~3.5% by weight)[4]. For different bird species, the eggshell exhibits a variety of sizes and thicknesses. Also, the shape of the eggshell varies dramatically across different species: conical in shorebirds, spherical in owls, and elliptical in hummingbirds[5]. At the nanoscale, the avian eggshell is composed of two layers of collagenized fibrous eggshell membranes and multiple calcified layers with different crystal orientations and nanostructures[6]. Extensive studies have been carried out to explain the structure-function relationship of eggshells. For instance, egg-shape variation could be attributed to the flight strength and efficiency of birds[5]. The well-organized multi-layered nanostructures of the eggshell change during the incubation, providing tunable mechanical properties to the eggshell and therefore weakening the eggshell for hatching[7].

Nacre and exoskeleton of lobster are also bioceramics that have the function of protecting the living organisms inside. Their extraordinary mechanical properties, such as high strength and toughness, originate from well-designed microstructures[8, 9]. For these bioceramics, the design objective is simple and straightforward: making the structure stronger and tougher. However, the eggshell has several design trade-offs that must be balanced. The eggshell must be sufficiently strong to resist the fracture caused by the load applied from the outside. Meanwhile, the eggshell must be easy to break from inside. In addition, the eggshell must be tough to delay the possible crack growth during the incubation, but the toughness must be weakened at the time of hatching. Moreover, the eggshell must be thin to save the calcium that is limited in the natural diet of birds. But the eggshell cannot be too thin to resist physical challenges from the environment. These trade-offs have already been balanced in the million years of evolution and the balance strategies are hidden in the sophisticated structures of the eggshell. However, little is known about the underlying mechanical design principles of the eggshell structure to balance the trade-offs. Here, we use experimental and computational approaches to elucidate such design principles in particular with regard to three key features of the eggshell structure: the dome-shaped shell, the membranes attached to the eggshell, and the exquisitely evolved thickness as shown in Figure 1A. More specifically, we will answer three questions: (i) Why the dome-shaped eggshell is hard to break from the outside but easy to break from the inside? (ii) How do membranes affect the toughness of the eggshell and make the toughness tunable? (iii) What is the possible driving force in the evolution of eggshell thickness? The general methodologies presented here hold great potential for the development of biologically inspired structural materials that can be used in sports safety applications and the packaging industry.

**RESULTS AND DISCUSSION**

Firstly, to access the breaking forces of the eggshell under the external load applied from the outside surface as well as the internal load applied from the inside surface, we performed two types of compression tests (Figure 2A and Figure 3A) on the half eggshell and compared the load-displacement curves. The typical load-displacement curves from two types of tests are shown in Figure 1B. The breaking force of the external compression test is much larger than the breaking force of the internal compression test. This large difference can be attributed to the two different fracture patterns induced by the dome-shaped structure. What are the two different fracture patterns and how do they affect the breaking force will be discussed later. Also, to evaluate the effect of membranes on the toughness of the eggshell, compression tests are performed on two types of eggshell samples: one with the membrane attached to the eggshell and the other one with the membrane removed. Two typical load-displacement curves from the two types of samples are shown in Figure 1C. The eggshell sample that has no membrane almost completely lost the load-carrying capacity after the breakage. However, the eggshell sample with the membrane maintains a certain amount of load-carrying capacity during the whole loading process. How the membranes dramatically increase the toughness of the eggshell will be discussed later. In the end, the data on the mass of eggs and thickness of eggshells from different species of birds are collected from previous research[10-12] and plotted in Figure 1D. The relationship between the shell thickness and mass tends to follow a special pattern (dashed line). In the following discussion, the underlying mechanical design principle of this special pattern will be uncovered and the possible driving force in the evolution of eggshell thickness will be presented.

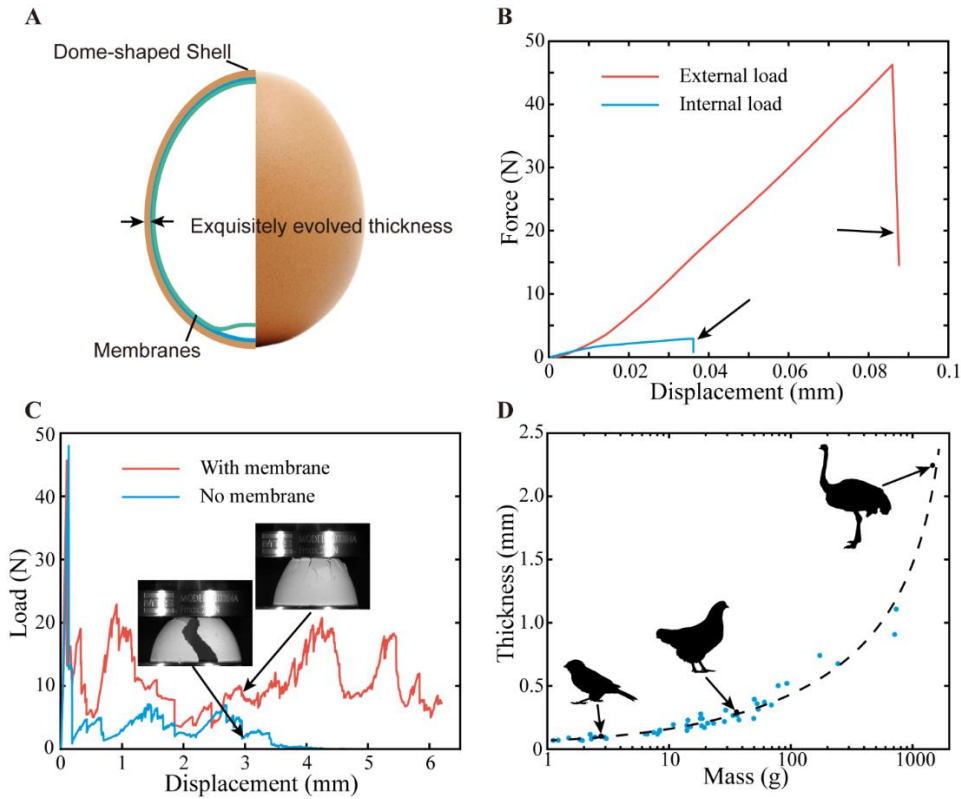

Figure 1. Structure and mechanical performance of eggshell.

(A) Three key features of the eggshell structure that balance the design trade-offs: the dome-shaped shell, the membranes attached to the eggshell, and the exquisitely evolved thickness. (B) The typical load-displacement curves of eggshells with internal and external load acting on them. The breaking force of the external compression test is much larger than the breaking force of the internal compression test. (C) Two typical load-displacement curves from eggshell samples with and without membranes. The eggshell sample with the membrane is much tougher than the other one. (D) Data points on the mass of eggs and thickness of eggshells from different species of birds. The relationship between the shell thickness and mass tends to follow a special pattern.

**Dome-shaped eggshell and two different fracture patterns**

Eggshells are made up of 95% calcium carbonate and can be considered a brittle material. For brittle material, in general, once the maximum principal stress exceeds the tensile strength, the small cracks initiate. Then these small cracks propagate rapidly and lead to a sudden catastrophic failure. Therefore, based on the theory of brittle fracture, a hypothesis can be made as follows. For eggshells in the compression tests, the cracks initiate at the position where the maximum principal stress ($\sigma_1$) reaches tensile strength, then the cracks propagate rapidly and lead to catastrophic failure. To verify this hypothesis, FE simulations are performed to obtain the stress distribution on the eggshell at the breaking point. Note that, the geometric parameters, material properties, and the breaking load applied to the FE model are measured experimentally. The detailed information can be found in the supplemental information.

For eggshells with internal compression, the cracked sample from the test is shown in Figure 2B. On the sample, 4 vertical cracks initiate from $\varphi = 90°$ and propagate to $\varphi \approx 80°$. The stress distribution results from the FE simulations are shown in Figures 2C and 2D. Note that, the stress varies through the thickness direction, therefore, the stress distributions on both the inside surface and outside surface are shown respectively. With the breaking internal load (2.6 N) applied to the model, the highest maximum principal stress can be found on the outside surface of the eggshell at $\varphi = 90°$. The value of the highest maximum principal stress at that point is 25.2 MPa which is slightly higher than the tensile strength (19.9 MPa) of the eggshell. According to the experimental and simulated results, the fracture follows the following process: with the internal load increasing gradually, maximum principal stress on the outside surface of the eggshell at $\varphi = 90°$ exceeds the tensile strength firstly, then, cracks at that position initiate and propagate rapidly and lead to a sudden catastrophic failure. Conclusively, the hypothesis we made before is verified by the observed fracture pattern in the test with internal load applied to the eggshell.

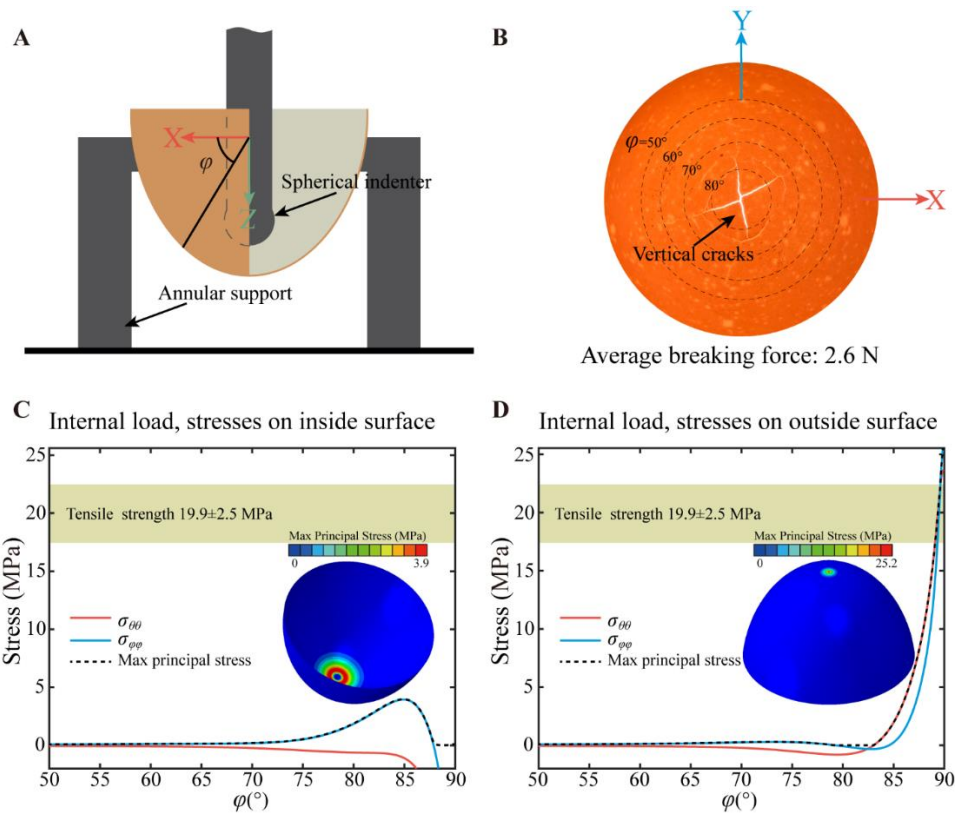

Figure 2. Internal compression tests.

(A) Schematic diagram of compression test with internal load. (B) Four vertical cracks initiate from $\varphi = 90°$ and propagate to $\varphi \approx 80°$ on the eggshell. (C) Stress distributions on the inside surface of the eggshell with breaking load applied on the FE model. Maximum principal stress is much lower than the tensile strength. (D) Stress distributions on the outside surface of the eggshell with breaking load applied on the FE model. Maximum principal stress is slightly higher than the tensile strength.

Similar experimental tests and simulations are performed on the eggshell with external load acting on it. On the eggshell sample, as shown in Figure 3B, a vertical crack initiates from $\varphi \approx 75°$ and propagates to $\varphi \approx 55°$. The stress distribution on the inside and outside surface from the FE simulations are shown in Figures 3C and 3D respectively. With the breaking external load (49.0 N) applied to the model, the highest maximum principal stress can be found on the inside surface of the eggshell at $\varphi = 90°$. According to the hypothesis proposed, there should be a crack initiated at this position which leads to the failure of the eggshell. However, both experimental and simulated results contradict the anticipation. Firstly, no visible cracks are found at $\varphi = 90°$. Secondly, the value of the highest maximum principal stress at that point is 480.0 MPa which is almost 24 times larger than the tensile strength (19.9 MPa) of the eggshell. If the crack initiated here leads to the failure, the breaking force would be much lower than the measured breaking force of 49.0 N. Three interesting findings from the results may provide a reasonable explanation for this paradox. Firstly, with the breaking internal load (49.0 N) applied to the model, the position that has the highest maximum principal stress on the outside surface is located at $\varphi \approx 80°$, while the crack initiation point observed from the test is at $\varphi \approx 75°$. Secondly, the value of the highest maximum principal stress at that point is 17.3 MPa which is just slightly lower than the tensile strength (19.9 MPa) of the eggshell. Lastly, at that point, the maximum principal stress $\sigma_1$ equals the stress component $\sigma_{\theta\theta}$ which leads to a vertical crack, while the crack observed from the test is vertical. All these findings lead to an assumption that two cracks initiate at different locations and different load magnitudes. One crack initiates on the inside surface at $\varphi = 90°$, and the load is way lower than the breaking load when its initiation. More importantly, this crack does not lead to catastrophic failure. The other crack initiates on the outside surface at $\varphi \approx 80°$, the load is the breaking load when its initiation. This crack propagates rapidly after initiation and leads to catastrophic failure. This assumption seems plausible, however, an important question needs to be answered: why the first crack does not lead to catastrophic failure?

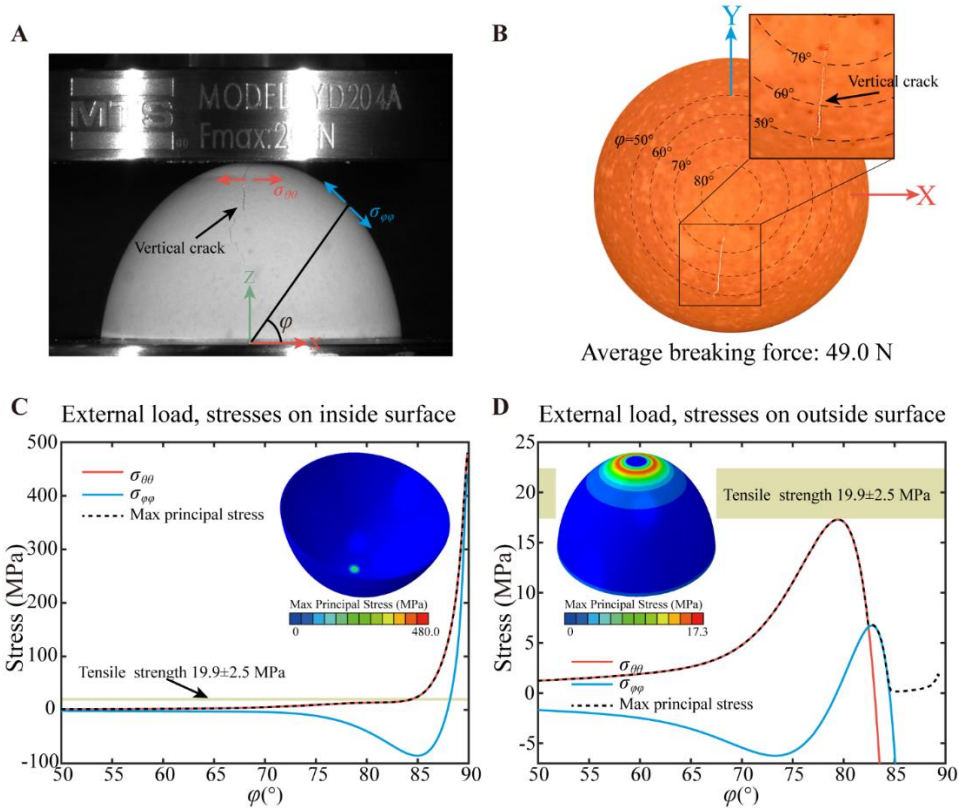

Figure 3. External compression tests.

(A) Schematic diagram of compression test with external load. (B) A vertical crack initiates from $\varphi \approx 75°$ and propagates to $\varphi \approx 55°$ on the eggshell. (C) Stress distributions on the inside surface of the eggshell with breaking load applied on the FE model. Maximum principal stress is almost 24 times higher than the tensile strength. (D) Stress distributions on the outside surface of the eggshell with breaking load applied on the FE model. Maximum principal stress is slightly lower than the tensile strength.

To answer the question, the detailed stress distribution in the through-thickness direction near $\varphi = 90°$ is obtained from the FE simulated results as shown in Figure 4A. The gray region which has the highest maximum principal stress is very narrow and surrounded by the lower maximum principal stress region. Especially in the thickness direction, when the crack propagates upward, it will almost immediately reach the compression-dominated region and could stop there. In fracture mechanics, the energetic stability determines whether a crack propagation is 'catastrophic' or 'controlled'. More specifically, if $dK/dC > 0$ the crack is stable, and if

$dK/dC < 0$, the crack is unstable and leads to catastrophic failure. Where $K$ is the stress intensity factor and $C$ is crack length. Therefore, a 3D FE model with a semi-elliptical surface crack on the inside surface is created to calculate the relationship between stress intensity factor and crack length as shown in Figures 4B and 4D. The stress intensity factors at two crack tips A and C are calculated to represent the whole crack front. Two crack aspect ratios are considered which are $a/c = 3$ and 6. The results are shown in Figures 4E and 4F, where all four curves follow the same 2-stage pattern. In the first stage, $dK/dC > 0$, the crack is small and unstable. The crack propagates rapidly until it reaches the peak of the curve and into the second stage. In the second stage, $dK/dC < 0$, stress intensity decreases as the crack length increases. The crack propagates slowly and stably and finally stops. In a previous study[13], a micro-crack (not detectable by routine visual inspection) at a load less than breaking force was found on the inside surface of the eggshell at $\varphi = 90°$ as shown in Figure 4C. This experimental result strongly supports the simulation result. To conclude, the eggshell with external load has a special two-crack fracture pattern. The first crack initiates on the inside surface at $\varphi = 90°$ and rests very soon while the crack is very small. The second crack initiates on the outside surface at $\varphi \approx 80°$. It propagates rapidly after initiation and leads to catastrophic failure. The rest of the first crack can be attributed to the dome-shaped structure-induced special stress distribution at the loading point.

Moreover, to further verify this conclusion, a validation test is performed as shown in Figure S4. The basic idea of the validation test is to apply geometric imperfections at the crack initiation positions to create stress concentration and accelerate the crack initiation. Two types of samples with geometric imperfections are prepared by drilling small holes at $\varphi = 90°$ (first crack) and $\varphi \approx 80°$ (second crack) respectively on the eggshell. For samples with a drilled hole at $\varphi = 90°$, the drilled hole does not affect the breaking force. However, the breaking force of samples with a drilled hole at $\varphi \approx 80°$ is much lower than the average breaking force of 49.0 N. This validation test further proves that the second crack leads to the failure of the eggshell.

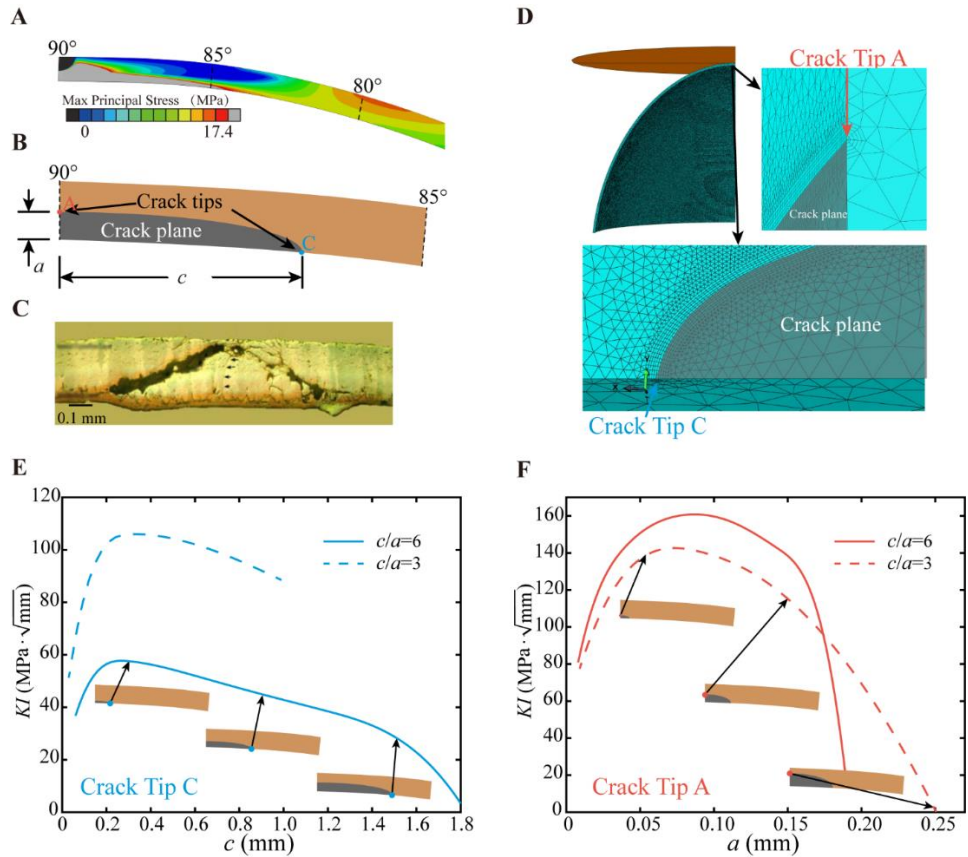

Figure 4. Crack resting mechanism.

(A) Maximum principal stress distribution in thickness direction near loading point. The gray region which has the highest maximum principal stress is very narrow and surrounded by a lower maximum principal stress region. (B) Schematic diagram of a semi-elliptic surface crack on the inside surface. Stress intensity factors will be calculated at crack tips A and C to determine if the crack is stable. (C) Micro-cracks near loading point at loads less than the breaking load from a previous study[13]. Experimental observations indicate this first crack does not lead to a catastrophic failure. (D) FE model and detailed mesh of a semi-elliptic surface crack on the inside surface. (E-F) FE simulated results of stress intensity factors at crack tips A and C respectively for a propagating crack. The second stage, $dK/dC < 0$, indicates that the crack propagates slowly and stably and finally stops.

Using the validated two-crack fracture pattern proposition, the breaking force of eggshell with load acting on different positions is predicted as shown in Figure S5. The predicted results match well with the experimental results. The results can also explain the reason why the eggshell is easier to break on the equator than on the two ends. Moreover, the effect of the two key geometric parameters, thickness, and curvature of the dome, on the breaking load of the thin-shell structure is systematically studied as shown in Figure S6.

**Membrane and the toughness of the eggshell**

The eggshell membrane is a collagenized fibrous tissue attached to the inside surface of eggshells. It has two layers of membranes: a thin inner membrane and a thicker outer membrane. The total thickness of the two layers of membranes is approximately 0.1mm[14, 15]. The eggshell membrane has been considered to have 3 essential functions for avian reproduction. It provides a structural foundation for the calcification of the eggshell[16]. It controls the gas exchange and helps the embryo breathe inside of the eggshell[17]. Also, it prevents bacterial penetration during the hatching process[18]. All these functions are based on the eggshell membrane's special structural, chemical, and physical properties, while the other hand, these properties have been utilized by many researchers in various fields. Eggshell membrane has been used as a template to fabricate hierarchically ordered microporous networks composed of oxide materials[19]. It also has been used as a polysulfide reservoir for highly reversible Li-S batteries[20]. However, the mechanical property of the eggshell membrane seems to be overlooked and few studies uncover the significant effect of eggshell membrane on the toughness of the eggshell.

To study the effect of membrane on the toughness of the eggshell, two groups of eggshell samples are prepared. For the first group of eggshells, the membranes are removed from the eggshell. The second group is the control group, and the membranes are still attached to the inside surface of the eggshell. The results of the two groups of samples from the compression tests are shown in Figure 5A. The breaking forces for the two groups of eggshells are nearly the same. The membrane has no significant effect on the strength of the eggshell considering it is so soft and thin compared to the eggshell.

Two distinctly different types of after-breakage force-displacement curves of the two groups of eggshell samples are shown in Figure 5B. Eggshells that have no membrane almost completely lose the load-carrying capacity after the breakage. However, eggshells with membranes maintain a certain amount of load-carrying capacity during the whole loading process. The initiation and propagation of cracks on the two groups of eggshells at different displacements are shown in Figures 5C and 5D. The effect of membrane on the toughness of eggshells can be summarized into two aspects.

Firstly, the main difference between the two types of samples is the propagation of the primary vertical crack. For eggshell with membrane, the primary vertical crack stops halfway to the bottom. While for the eggshell has no membrane, the primary vertical crack propagates all the way to the bottom (Figure 5C, 5D and Figure S7). Just like organic matrix bridging in nacre[21] and collagen-fibril bridging in bone[22]. Eggshell membrane can bridge a developing crack and carry the load that would drive the propagation of the crack. It lowers the stress intensity factor at the vertical crack tip and stops the propagation of the primary vertical crack.

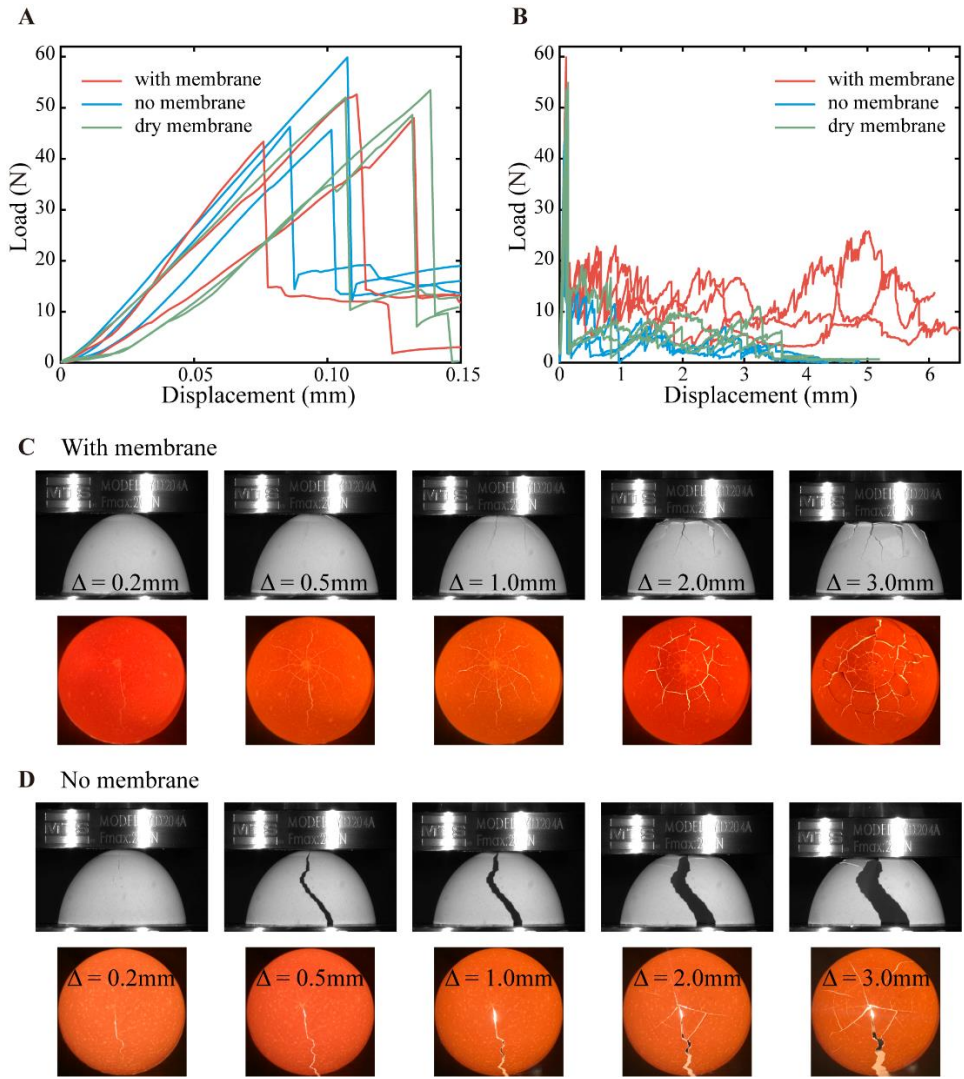

Figure 5. Membrane and the toughness of the eggshell

(A) Pre-breakage force-displacement curves of three types of eggshells: membrane attached, membrane removed, and dry membrane attached. The membrane has no significant effect on the strength of the eggshell considering it is so soft and thin compared to the eggshell. (B) Post-breakage force-displacement curves of three types of eggshells. Eggshells that have no membrane or dry membrane almost completely lose the load-carrying capacity after the breakage. However, eggshells with wet membranes maintain a certain amount of load-carrying capacity during the whole loading process. (C-D) Crack initiation and propagation of eggshell with and without membrane attached. For eggshell with membrane, the primary vertical crack stops halfway to the bottom. While for eggshell that has no membrane, the primary vertical crack propagates all the way to the bottom. Because of the membrane, the high stress that cannot be effectively relieved by the primary crack, the membrane changes the stress distribution and creates many secondary cracks that improve the toughness and energy absorption ability of the eggshell.

Secondly, because the membrane stops the propagation of primary vertical cracks, the overall stress distribution and secondary cracks are changed accordingly. After the primary vertical crack, multiple vertical and circumferential secondary cracks alternately initiate and propagate on the eggshell. These two types of cracks are driving by two different types of stress components: $\sigma_{\theta\theta}$ that controls the vertical crack and $\sigma_{\varphi\varphi}$ that controls the circumferential crack. These two stress components are competing during the entire loading process. The membrane dramatically affects the competing results as shown in Figures 6A and 6B. For eggshell with membrane, the three drops in the force-displacement curves represent three different crack initiations. From points 1 to 3, the winners of the competition are $\sigma_{\theta\theta}$, $\sigma_{\varphi\varphi}$, and $\sigma_{\theta\theta}$. Therefore, the three cracks initiated sequentially are vertical crack, circumferential crack, and vertical crack. However, for the eggshell that has no membrane, the stress distributions are completely different after the initiation of the first vertical crack. Therefore, the three cracks initiated sequentially are vertical crack, vertical crack, and circumferential crack. Moreover, the simulated results also show that eggshell without membrane has much lower load-carrying capacity.

Except for the static compression test, dynamic impact tests are also performed on these two types of eggshells as shown in Figures S8A, S8B. An impactor with a speed of 2.7m/s hits the eggshell, and this impact process is captured by a high-speed camera. The whole impact process lasts for about 6ms. The eggshell with the membrane remains in one piece after the impact while the eggshell has no membrane breaks into many pieces. Using the recorded pictures, the velocity change of the impactor, as well as the reaction force-displacement curve, are calculated and shown in Figures S8C and S8D. Clearly, the eggshell with the membrane absorbs more impact energy. In conclusion, because of the membrane, the high stress that cannot be effectively relieved by the primary crack, the membrane changes the stress distribution and creates many secondary cracks that improve the toughness and energy absorption ability.

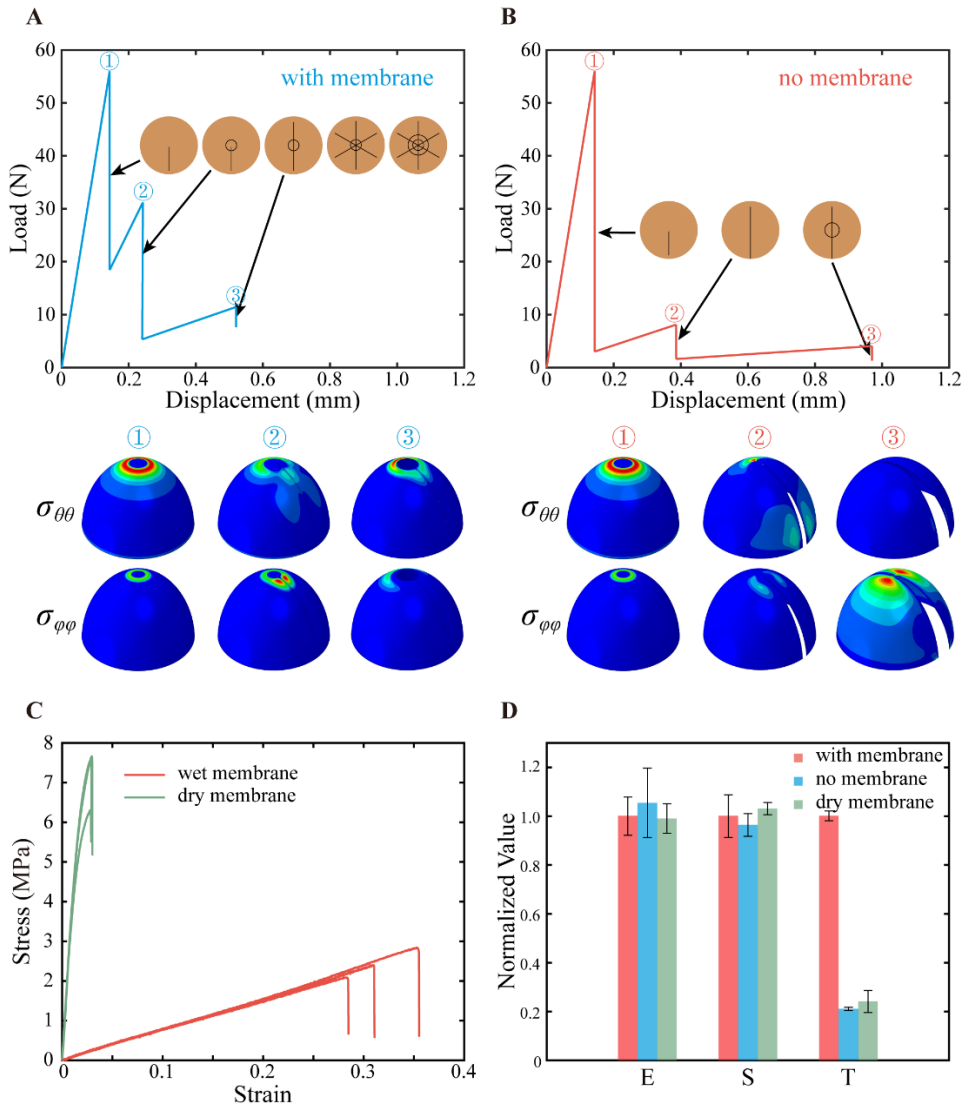

Figure 6. Competition between vertical and circumferential crack propagation.

(A-B) Crack propagation behaviors of eggshells with and without membrane. Multiple vertical and circumferential secondary cracks alternately initiate and propagate which are driving by $\sigma_{\theta\theta}$ $\sigma_{\varphi\varphi}$ respectively. The membrane dramatically affects the stress distribution on the eggshell during the crack propagation and therefore affects the competition between vertical and circumferential crack propagation. Because of that, eggshell without membrane has a much lower load-carrying capacity. (C) Stress-strain curves of dry and wet eggshell membranes. Moisture content can dramatically affect the mechanical property of the membrane. The failure strain of the dry membrane is about 0.03 but the failure strain of the wet membrane is about 0.3, 10 times higher than the dry membrane. (D) Three typical mechanical properties (stiffness, strength, and toughness) comparison between the three groups of eggshells (wet membrane, dry membrane, and no membrane).

The membrane makes the eggshell tougher and protects the embryo inside, but it also leads to a problem: does the membrane impede the hatching process and make it much hard for the chick to break out from the eggshell? Previous studies revealed many physical and biological transitions of the eggshell during the incubation process that will help the chick to break the eggshell. For example, the incubated eggshell is 7.3% thinner than the unincubated eggshell[23], and the pore density of the eggshell increases by about 12%[24]. During the incubation, avian eggs lose approximately 20% water, and the membranes dry out accordingly[25]. However, no one realized that the change of moisture content of the membrane is also an evolutionary strategy that makes the eggshell weaker during the hatching process. To uncover the effect of the moisture content, another group of eggshell specimens are created and dried in the air for 24 hours and then tested. The results of compression tests are shown in Figures 5A and 5B. The eggshell with the dry membrane has the same breaking force as the eggshell with the wet membrane. However, the post-breaking force-displacement curves show that the toughness of the dry/wet membrane groups is totally different. The force-displacement curve of the dry membrane group is more like that of the group has no membrane attached to the eggshell. The fracture patterns (Figure S9B) are also very similar to the group of eggshells have no membrane.

Why is the toughness of the eggshells with dry and wet membranes distinctly different? That is because the moisture content can dramatically affect the mechanical property of the membrane. Dry and wet membrane specimens (Figure S9A) are prepared and tested; the stress-strain curves are shown in Figure 6C. The failure strain of the dry membrane is about 0.03 but the failure strain of the wet membrane is about 0.3, 10 times higher than the dry membrane.

Therefore, when the eggshell breaks, the dry membrane breaks too, but the wet membrane does not break with the eggshell, instead, it provides a closure force along the crack path and stops the crack propagation.

Three typical mechanical properties of the three groups of eggshells: stiffness ($E$, slope of the force-displacement curve), strength ($S$, the breaking force), toughness ($T$, the area under the static force-displacement curve) are calculated and compared in Figure 6D. The results indicate that the wet eggshell membrane with a large breaking strain greatly increases the toughness of the eggshell. During the incubation period, the moisture content of the membrane decreases, and the breaking strain decreases correspondingly which leads to a dramatic decrease of the toughness of the eggshell and helps the chick to break out of the eggshell easily.

**A possible driving force in the evolution of eggshell thickness**

For different species of birds, the mass of the eggs and the thickness of the eggshells are different. The mass of the ostrich (*Struthio camelus*) egg is about 1500g and the thickness is around 2 mm[26]. Where the mass and thickness of chicken (*gallus domesticus*) egg are about 50g and 0.38mm respectively[27]. Why the eggshell of chicken is not thicker, i.e., 0.6mm, or thinner, i.e., 0.2mm? From an evolutionary standpoint, the process of evolution is the survival of the fittest, so 0.38mm could be the best thickness in natural selection. Many efforts have been made trying to explain such natural selection. A general explanation is that the eggshell is created to satisfy opposing goals. It must be thick enough to withstand the external forces imposed on it during the incubation. On the other hand, it cannot be too thick to waste the building material or too thick to inhibit the hatchling to break its way out[10, 28]. To uncover what parameters that may affect the eggshell thickness, further studies have been carried out. These parameters including the egg mass[29], incubator mass[30], breeder age[31]. These studies provided some regression equations to describe the relationship between those factors and eggshell thickness. However, these studies summarized the phenomenon observed in many experiments, but still, they didn't touch the essence of what is the driving force in the evolution of eggshell thickness.

The thickness of the eggshell seems to be controlled by a balance between protection and nutrient consumption, to maximize survivability. Reproduction is energetically costly and requires a sufficient supply of macronutrients and micronutrients. Among these, calcium is probably the most limiting micronutrient[32]. To quantitatively study the contradictory relations between the protection and calcium consumption, a three-index model is created as shown in Figure 7A. The first one is the production index, *PI*, which is defined as:

$$PI(t) = \frac{V_{egg}}{V_{eggshell}(t)} \quad (1)$$

Where *t* is the thickness of the eggshell; $V_{egg}$ the volume of the egg which is considered as the limited calcium that can be used to produce eggshell; and $V_{eggshell}$ is the volume of the eggshell itself and it is a function of thickness. The thicker the eggshell is, the fewer eggs can be produced by the limited building material. For example, as shown in Figure 7A, with a small thickness, $t_s$, 20 eggs are produced, but with a larger thickness, $t_l$, only 5 eggs are produced. More eggs produced doesn't necessarily mean that more offspring can survive because eggs are fragile, and they could be broken under environmental loads. So here, the second parameter is introduced, *BI*, breakage index which defines how many eggs break under the environmental loads. In our example, for eggs with small thickness, $BI(t_s) = 14$, which means 14 eggs break under the environmental load. The last parameter is the survival index,

$SI = PI - BI$, which indicates how many eggs survive. Among all three thicknesses, medium thickness, $t_m$, has the highest survival index and wins the natural selection. In conclusion, our hypothetical model assumes that the driving force in the evolution of eggshell thickness is to maximize the survival index. Therefore, the optimal thickness of eggshells is the thickness that has the maximum survival index. To validate this hypothetical model, the relationship between $SI$ and eggshell thickness needs $t$ to be uncovered.

Considering $SI$ is the difference between $PI$ and $BI$, and $PI$ can be easily obtained using Equation 1, determining $BI$ is the key to finding the optimal thickness. $BI$ can be calculated with the following 3 steps. The first step is to determine the relationship between the breaking force, $F$, and thickness, $t$, of the eggshells. In the previous sections, we demonstrated that when the highest maximum principal stress on the outside surface of the eggshell reaches the tensile strength, the corresponding load is the breaking force of the eggshell. Based on that, FE models of eggshells with different thickness are created and their breaking forces are obtained from the simulation results which is shown in Figure 7B.

After having the function $F(t)$ that describes the relationship between breaking force and thickness. The second step is to calculate the breaking possibility using $F(t)$. Here we assume the maximum environmental loads of different eggs follow a normal distribution which can be described using the following equation:

$$y = \frac{1}{\sigma\sqrt{2\pi}} e^{\frac{-(x-\mu)^2}{2\sigma^2}} \tag{2}$$

Where $y$ is the probability density, $x$ is the maximum environmental load, $\sigma$ and $\mu$ can be calculated using the following equation:

$$\sigma = \frac{m \cdot g \cdot k}{x_0 + 3} \quad , \quad \mu = 3\sigma \tag{3}$$

Where *m* is the mass of the egg, *g* is the gravitational acceleration, and *k* is a load factor which is set as 100 here. $x_0$ can be solved using the following equation:

$$E = \frac{1}{\sigma_0 \sqrt{2\pi}} \int_{-\infty}^{x_0} e^{\frac{-(t-\mu_0)^2}{2\sigma_0^2}} dt \tag{4}$$

where $\mu_0 = 0$, $\sigma_0 = 1$. *E* is an environmental factor that controls the shape of the probability density function. Three different probability density functions with *E*=0.5, 0.7 and 0.9 are shown in Figure 7C. A smaller *E* indicates a much harsher environmental condition and the probability of a large environmental load acting on the egg is higher. With this given force, the breaking probability of the eggshell under environmental load can be calculated with the following equation:

$$p = \frac{1}{\sigma \sqrt{2\pi}} \int_{F}^{\infty} e^{\frac{-(t-\mu)^2}{2\sigma^2}} dt \tag{5}$$

For instance, assume an eggshell has a breaking force of 60 N in a mild environment, *E* = 0.9, the value of breaking probability *p* is the area of the green region. In a harsh environment, *E* = 0.5, the value of breaking probability *p* is the area of the green, blue and red regions together. To this point, the breaking index *BI* can finally be calculated:

$$BI = PI \cdot p \tag{6}$$

With *E* = 0.8, the three indexes are calculated for eggshells with different thicknesses from 0.2 mm to 1.0 mm as shown in Figure 7D. Production index and breaking index both decrease as the thickness increases, and the breaking index decreases much faster. The survival index increases first and then decreases, which creates a peak point. The thickness corresponding to the peak value of the survival index is $t \approx 0.4$ mm which is very close to the average thickness of eggshell, $t \approx 0.38$mm, observed from the tests.

To further validate the hypothetical model, survival indexes for different environmental factors are calculated and the results are shown in Figure 7E. The red line connects all highest survival index for different environmental factors is the corresponding optimal thickness. The optimal thickness is located in a very narrow range, from 0.34mm to 0.44mm. Also, the optimal thickness is thinner for mild environmental conditions and thicker for harsh environmental conditions which is reasonable.

All the discussion above is focused on chicken eggshells, but the hypothetical model itself should also be able to apply to other species of birds. Therefore, the relationship between optimal thickness and the mass of the egg is calculated using the discussed calculation procedure. The gray region in Figure 7F represents the optimal thickness for different environmental factors from 0.5 to 0.99. Evidently, the majority of the observed data points are in or very near to the gray region which strongly validates the 3-index hypothetical model. To conclude, the possible driving force in the evolution of eggshell thickness is the balance between protection and calcium consumption, to maximize survivability

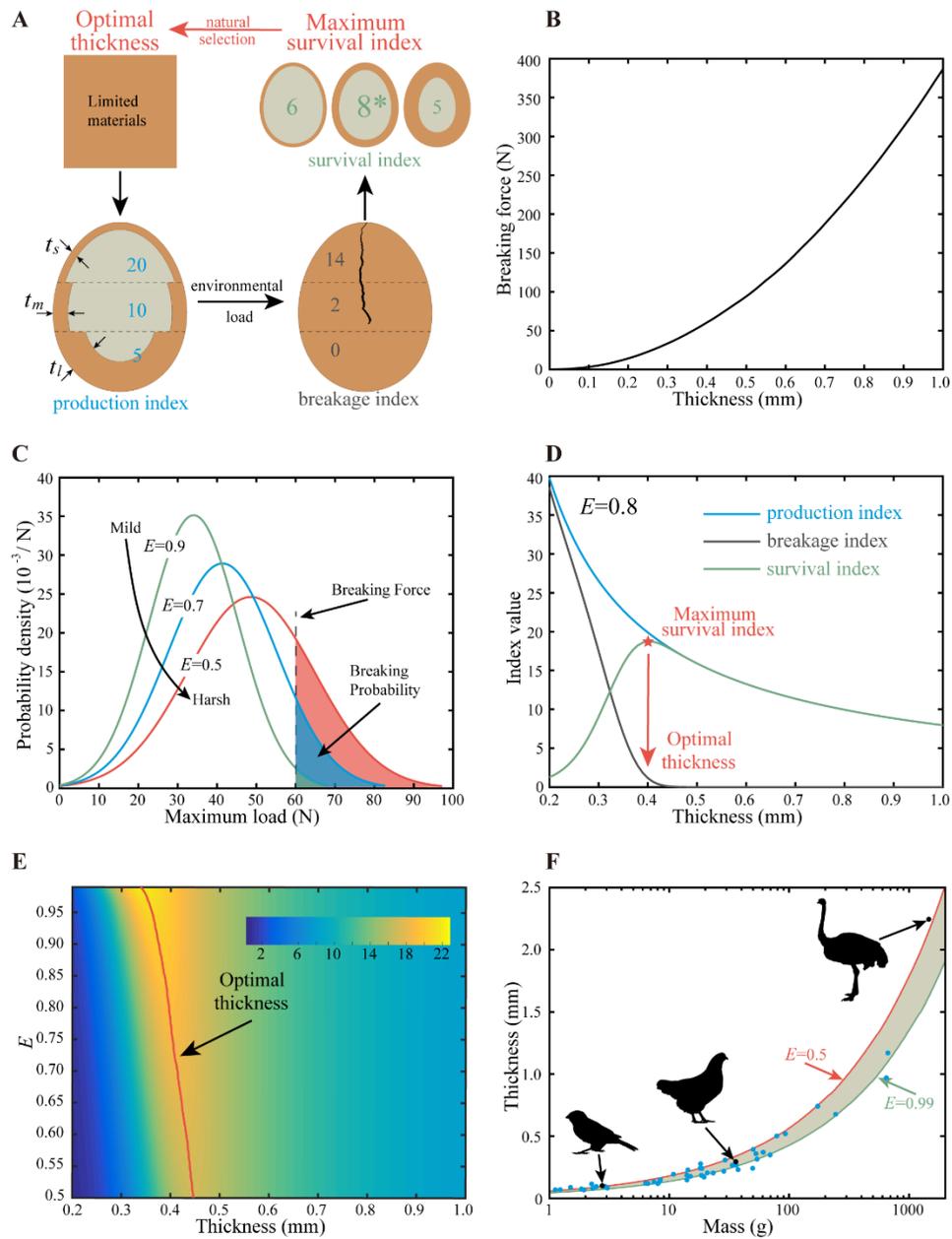

Figure 7. A possible driving force in the evolution of eggshell thickness. (A) Schematic diagram of the three-index model. The thickness of the eggshell seems to be controlled by a balance between protection and nutrient consumption, to maximize survivability. (B) Breaking forces of eggshell with different thickness. (C) Distribution of maximum environmental load of eggshell. $E$ is an environmental factor that controls the shape of the probability density function. A smaller $E$ indicates a much harsher environmental condition and the probability of a large environmental load acting on the egg is higher. (D)The production index, breakage index, and survival index for eggshells with different thicknesses. The thickness corresponding to the peak value of the survival index is $t \approx 0.4$ mm which is very close to the average thickness of eggshell, $t \approx 0.38$mm, observed from the tests. (E) Survival index of eggshells with different environmental factors and thickness. The optimal thickness is located in a very narrow range, from 0.34mm to 0.44mm. (F) Calculated optimal thickness of eggshell for different bird species. The majority of the observed data points are in or very near to the calculated gray region which strongly validates the proposed 3-index hypothetical model.

**CONCLUSION**

Due to its special function, the eggshell has three main trade-offs in the reproduction which are: 1) the eggshell should be easy to break from the inside but hard to break from the outside; 2) the eggshell should be tough during the incubation but be weakened at the time of hatching; 3) the eggshell should be thin to save the building material but not too thin to be easily broken. In this paper, we have investigated the structure of the eggshell and the underlying mechanism that balance the trade-offs. More specifically, we studied the crack initiation and propagation of the eggshell under two different types of loads, the internal and external loads. The initiation positions for the two types of loads are the same, at the opposite surface of the load points. For internal load, the crack propagates after initiation and then causes the failure of the eggshell. However, for external load, the crack stops after initiation because the surrounding area is compression-dominated. Then, a second crack initiates on the outside surface propagates fast, and causes the catastrophic failure of the eggshell. These two distinct fracture patterns account for the large difference in breaking forces between the external and internal load conditions. Also, we studied the effect of the membrane on the toughness of the eggshell and confirmed that the wet membrane can dramatically increase the toughness of the eggshell. While the dry membrane has a much lower breaking strain and doesn't contribute to the toughness. During the incubation, the moisture content of the membrane declines continuously which makes the eggshell much weaker at the time of hatching. Lastly, we investigated the relationship between eggshell thickness and breeding efficiency by introducing a 3-index model. The theoretical optimal thickness of the eggshell was calculated and compared with the collected thickness data of different bird species. The comparison results verified the rationality of the 3-index model and concluded that the driving force in the evolution of eggshell thickness is to balance the production index (PI) and breakage index (BI) and archive the highest breeding efficiency. The general methodologies presented here hold great potential for the development of eggshell-inspired structural materials that can be used in sports safety applications and the packaging industry.

**EXPERIMENTAL PROCEDURES**

**Specimen Preparation**

The eggs were obtained from a retail outlet, described as "Large Grade AA Cage Free Organic eggs". The external dimensions ($a$, $b$, $e$) of the eggshell were measured before testing and the thickness $t$ was measured after testing.

To obtain the compression specimens, the eggs were cut in half using a hand drill with a diamond saw rotating disc. The cutting edge was then sanded with 1000 Grit Sandpaper. The half eggshell specimens were washed with distilled water and checked under the light. Specimens with any visible cracks were discarded. Three types of specimens were prepared. Specimens with wet membranes attached to them were stored in water. Specimens with dry membranes were dried in air for 24 hours. For specimens without membranes, the membranes were removed using clamps.

To obtain the membrane specimens for tensile tests, the eggshells were cut in half first. Then the eggshells were rinsed in 1 M HCl to dissolve CaCO3, leaving the organic shell membranes. After the shell membranes were thoroughly washed with water, the dogbone shape specimens were cut out of the membrane in the latitudinal direction. Half of the specimens were stored in water and the other half were dried in air for 24 hours.

**Mechanical Testing**

To capture the mechanical response of the eggshell and membrane, compression test and uniaxial tensile tests were performed using an MTS mechanical tester (C43) with 1kN and 100N load cells. Both the compression test and tensile test were conducted in a quasi-static regime. The displacement rate for the compression test was 0.5 mm/min and the strain rate for the tensile test was 0.001/sec. The toughness was measured from the load-displacement which is defined as the area under the force-displacement curve which represents the energy absorbed and dissipated. This definition is wildly used in rigid biological composites. Images of the specimens during the loading procedures were captured at a rate of 1 FPS.

The impact tests were performed on self-assembled impact test equipment as shown in Figure S10. The impactor moves downwards in the guide rail. Before contacting the eggshell, the trigger activates the high-speed camera's shutter. The impact duration is then recorded by a high-speed videoing system (Photron SA1.1) with 20000 frames per second. Displacement and velocity histories are then calculated by digital image correlation (DIC) using the commercial software VIC-2D (Correlated Solution Inc.).

**Finite-Element Simulations**

Numerical simulations are conducted using the commercial FE package ABAQUS/Standard to capture the stress distribution on the eggshell for different compression and tensile tests. The model of the eggshell structure is generated using 8-node linear brick, reduced integration, hourglass control solid element C3D8R with an elastic model (Young's modulus of 30.0 GPa and Poisson's ratio of 0.3) to define the material property. To get accurate stress distributions in the thickness direction, the mesh size in the thickness direction is set to $t/10$ and the mesh size is verified by a mesh sensitivity test. The indenters are modeled with a 3D analytic rigid shell. Contact effects are modeled using a hard contact behavior for the normal direction.

Another type of FE simulation is performed to get the stress intensity factors of the cracks. The square root singularity of stress at the crack front region is modeled by shifting the mid-point nodes to the quarter-point locations. This area meshes with collapsed elements C3D15 and the rest of the model is covered with C3D10 elements. the size of the elements is small enough near the crack front to ensure the accuracy of the results as shown in Figure 4D. The stress intensity factors are calculated from the ABAQUS built-in J-integral solver based on the domain integral technique. The reliability of the finite element model and the accuracy of the calculated stress intensity are checked by performing the convergence tests.

**ACKNOWLEDGMENTS**

The author would like to thank Professor Toshio Nakamura for his helpful advice on various technical issues in this Paper.

**AUTHOR CONTRIBUTIONS**

## DECLARATION OF INTERESTS

The authors declare no competing financial interests.

**Supplemental Information**

**Mechanical design principles of avian eggshells for Survivability**

Fan Liu, Xihang Jiang, and Lifeng Wang

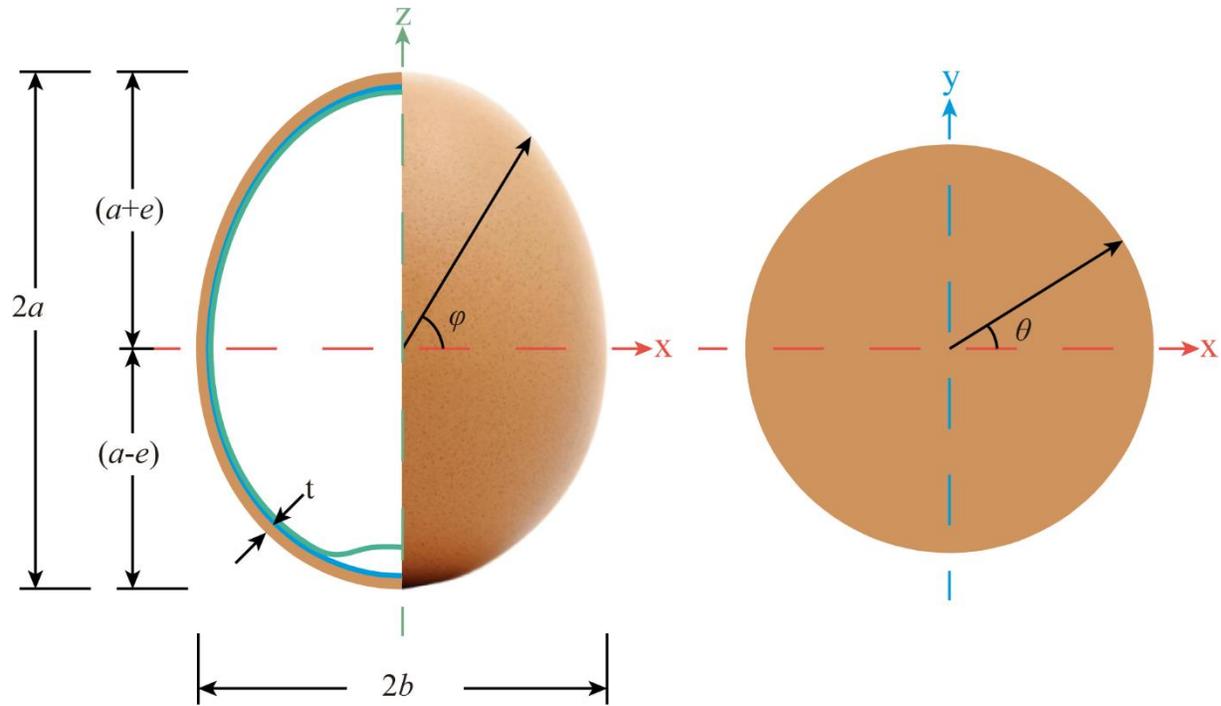

Figure S1. Diagram defining the terms in the equation for the egg profile. The profile of the egg is given by the equations: $x = b\cos\alpha$ and $y = (a + e\sin\alpha)\sin\alpha$. In this study, three parameters ($a$, $b$, and $e$) are measured from the eggshell samples and their average values are used in the FE models. The average values are $2a = 56.4$ mm, $2b = 44.6$ mm, and $e = 2.1$ mm.

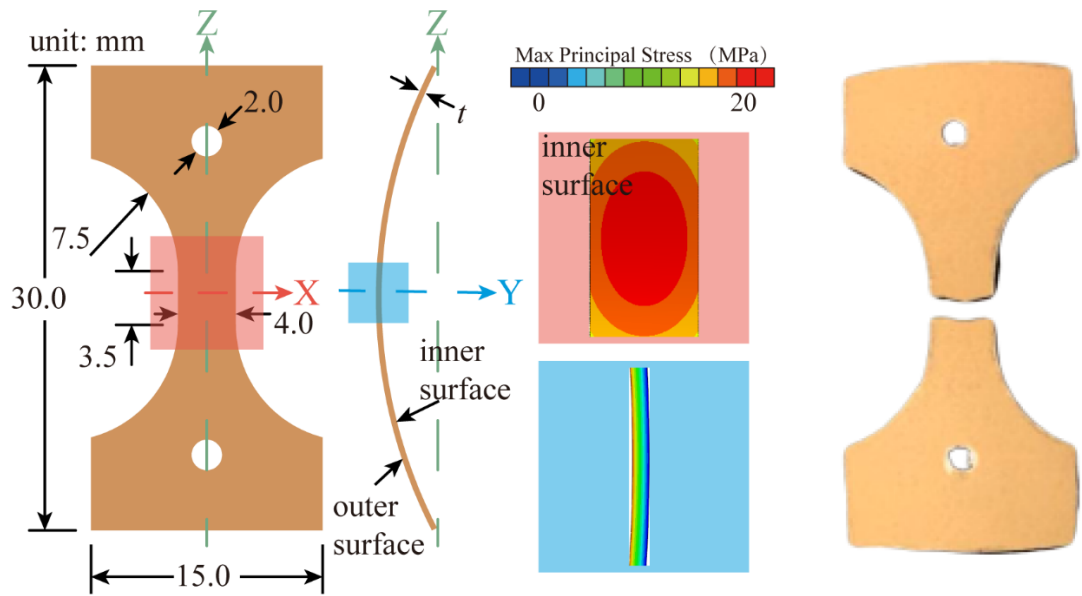

Figure S2. The geometric parameters of the tensile specimen of the eggshell that used for obtaining the tensile strength. Note that, unlike the regular tensile specimen which is flat, the specimen used here is curved. The curvature shape makes the stress distribution more complex and therefore the tensile strength cannot be calculated directly from the experimental test results. Therefore, a FE model of the curved tensile specimen is created and the result of the tensile strength is obtained from both the breaking load from the experimental test and the stress distribution from the FE simulation.

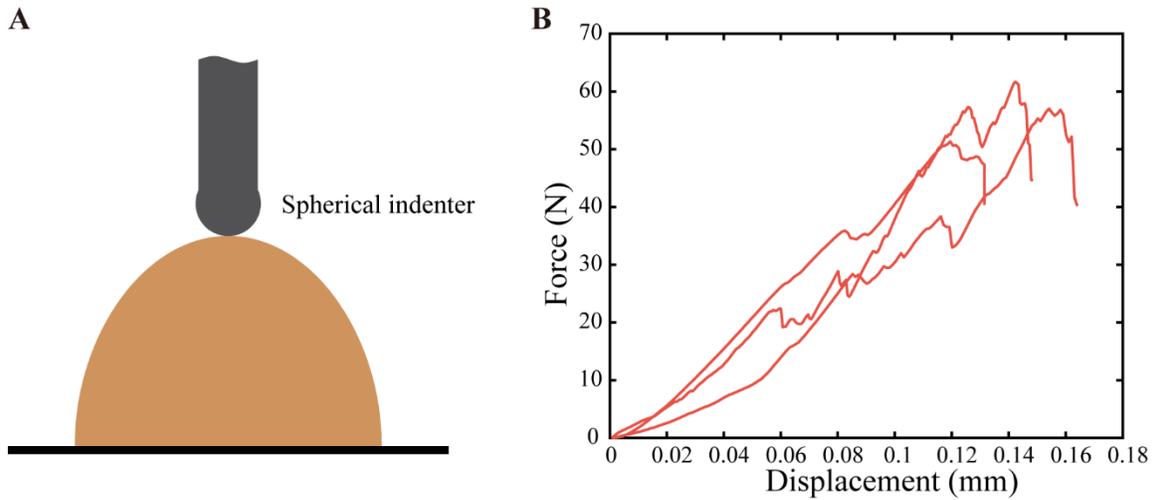

Figure S3. Extra external compression test with a spherical indenter. The two types of compression tests described in the main manuscript used indenters with different shapes: spherical indenter for internal compression test and flat indenter for external compression test. It is reasonable to doubt that the shape of the indenter may affect the breaking force. However, the breaking force obtained from the external compression tests with a spherical indenter is almost the same as the breaking force obtained from the external compression tests with a flat indenter in the main manuscript. So, the shape of the indenter does not have an apparent effect on the test results of breaking force.

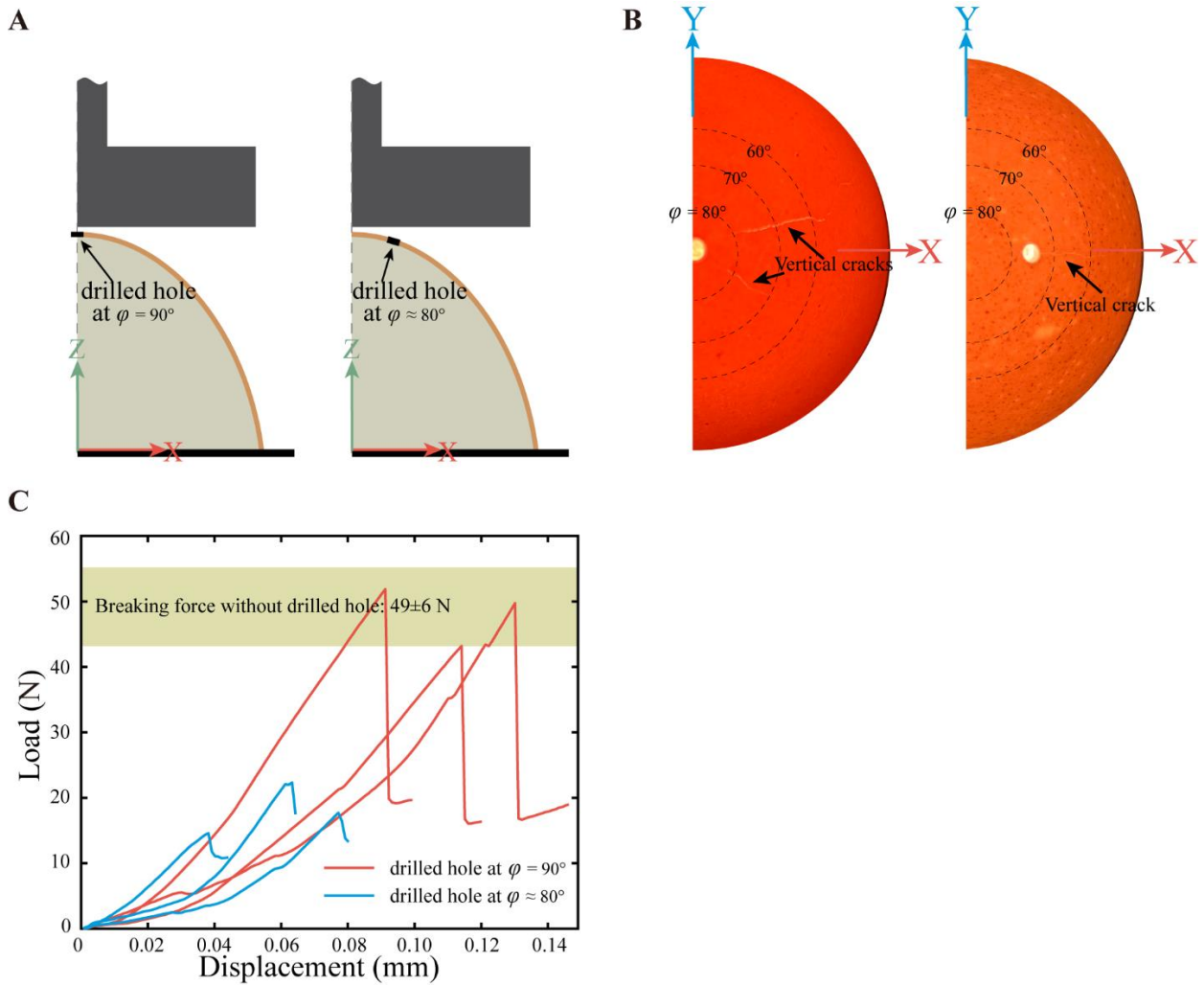

Figure S4. (A-B) The external compression test with pre-drilled holes at two crack initiation positions. Two types of samples with geometric imperfections are prepared by drilling small holes at $\varphi = 90°$ (first crack) and $\varphi \approx 80°$ (second crack) respectively on the eggshell. (C) For samples with a drilled hole at $\varphi = 90°$, the drilled hole does not affect the breaking force. However, the breaking force of samples with a drilled hole at $\varphi \approx 80°$ is much lower than the average breaking force of 49.0 N. This validation test further proves that the second crack leads to the failure of the eggshell.

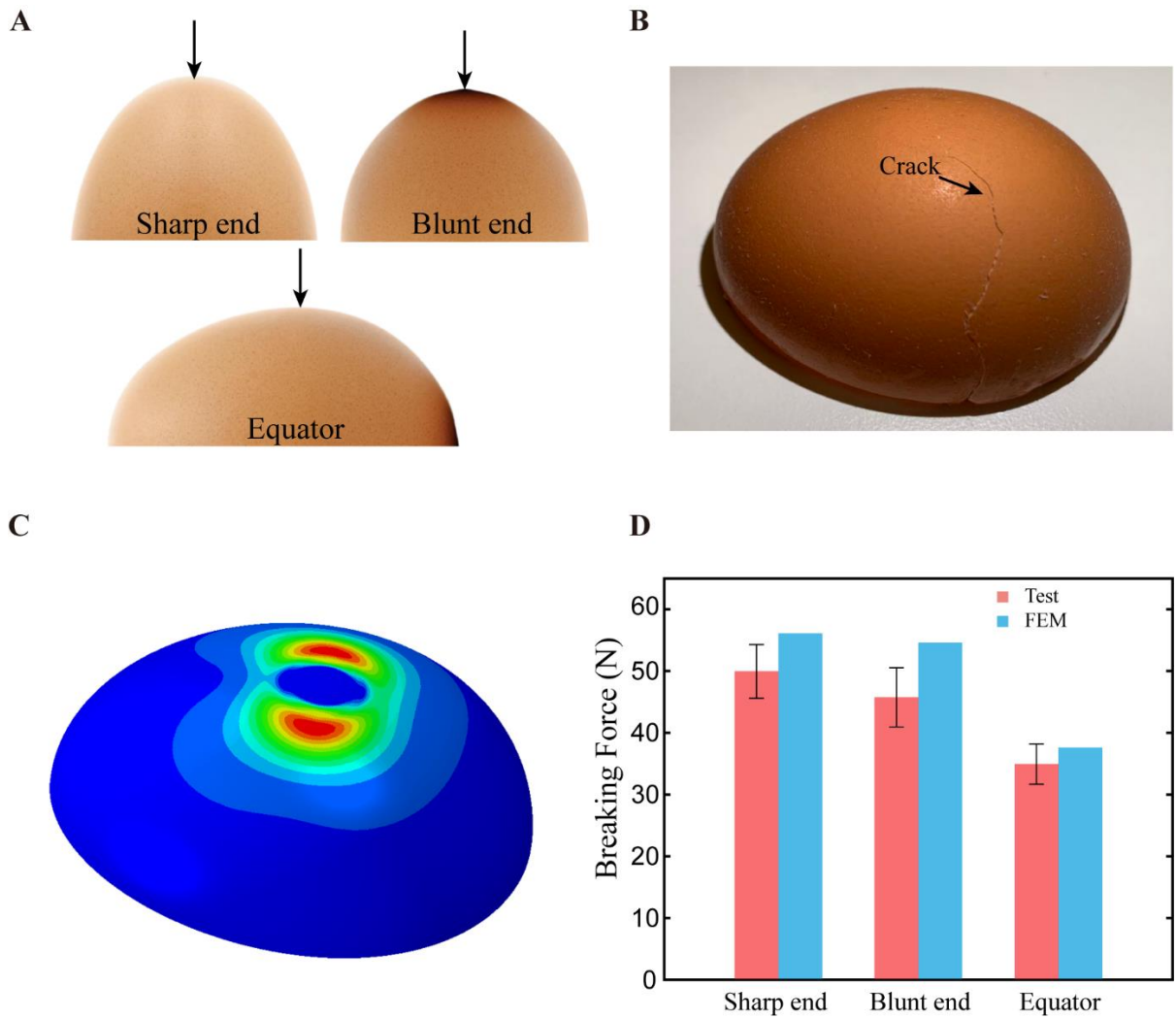

Figure S5. (A) The external compression tests are performed on three typical locations of the eggshell: sharp end, blunt end, and the equator. (B-C) With external load applied to the equator, the crack propagates vertically along the equator which matches well with the FE simulated principal stress distribution on the outside surface of the eggshell. (D) Moreover, using the validated two-crack fracture pattern proposition, the breaking force of eggshell with load acting on different positions is predicted. The results match well with the experimental results. Because of the different curvatures, stress distribution varies at different locations, and therefore breaking force varies. The eggshell is easier to break on the equator for its lower curvature.

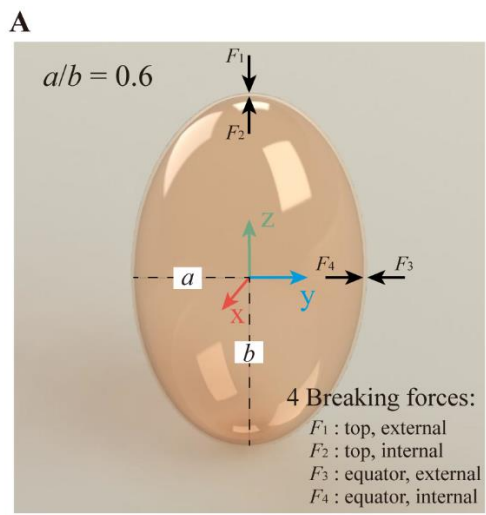
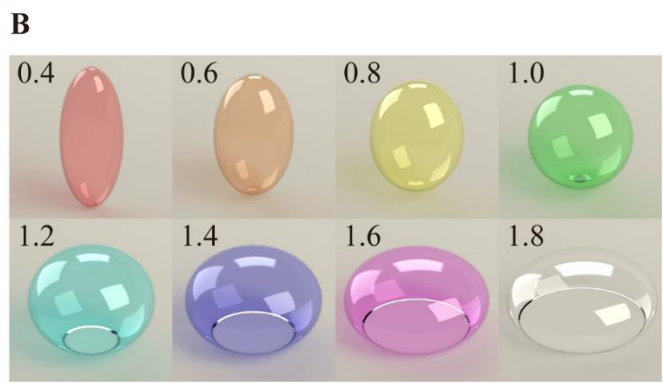
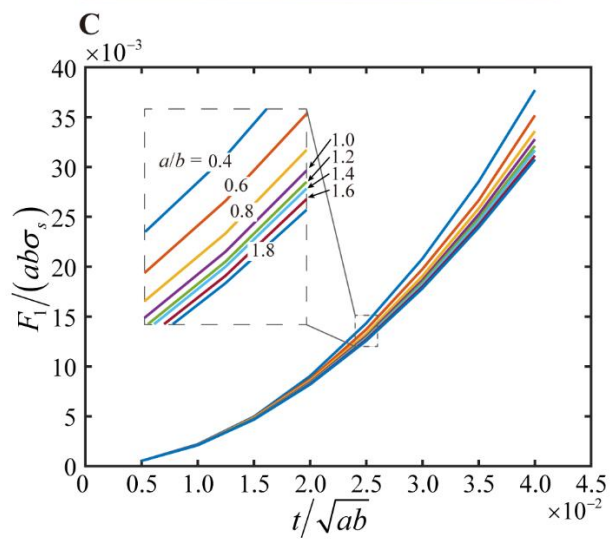
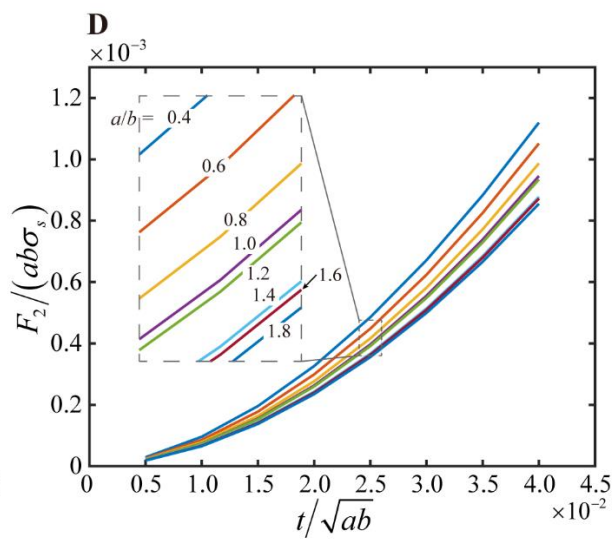
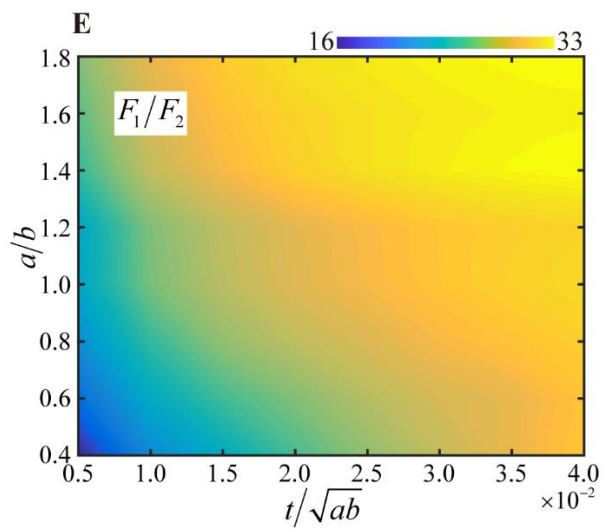
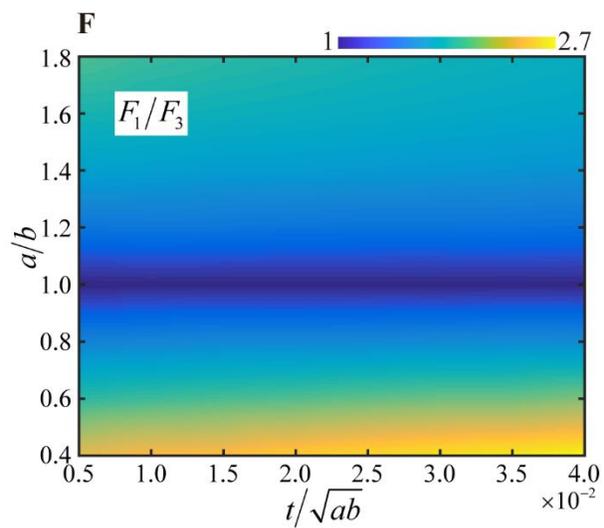

Figure S6. To study how the two important geometric parameters (thickness and curvature) affect the breaking force of egg-like thin shell structures. (A-B) FE models with different $t$ and $a/b$ are created. Note that the shape of these models can be described by the equations shown in Figure S1 but $e$ is set to 0 for all models. Four different breaking forces are calculated in the FE simulation. $F_1$ is the external breaking force at the top. $F_2$ is the internal breaking force at the top. $F_3$ is the external breaking force at the equator. $F_4$ is the internal breaking force at the equator. (C-D) Normalized top breaking forces of FE models with different $t$ and $a/b$ are calculated. Both external and internal breaking forces increase as the thickness increases. Also, for models with the same thickness, the smaller $a/b$, the larger the internal and external breaking force. (E) The comparison between the internal and external breaking force applied on the top of the shell structure. The external breaking force is much larger than the internal breaking force for all models. Also, the difference increase as the thickness increases or the $a/b$ increases. (F) The comparison between the external breaking forces applied on the top and the equator of the shell structure. The thickness has little effect on the $F_1/F_3$ while the $a/b$ do affect the $F_1/F_3$. When $a/b = 1$, the structure is a spherical shell, naturally, $F_1/F_3 = 1$. However, for models with $a/b > 1$ or $a/b > 1$ breaking force at the top is always larger than the breaking force at the equator.

**A** With membrane

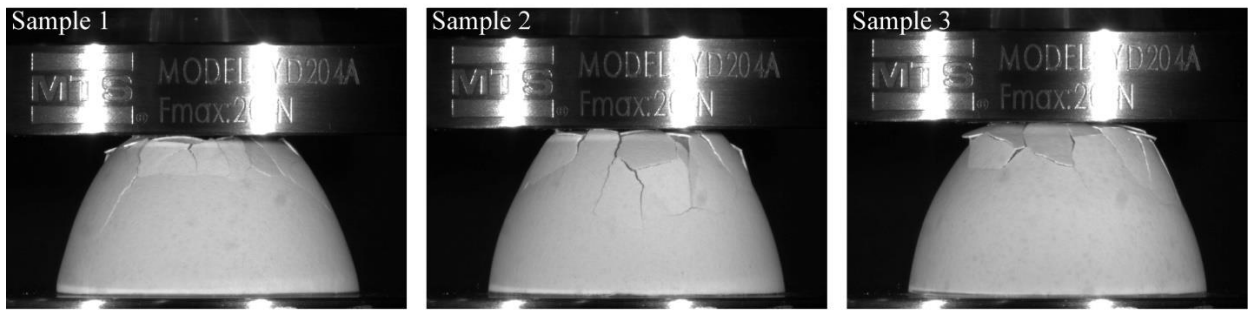

**B** No membrane

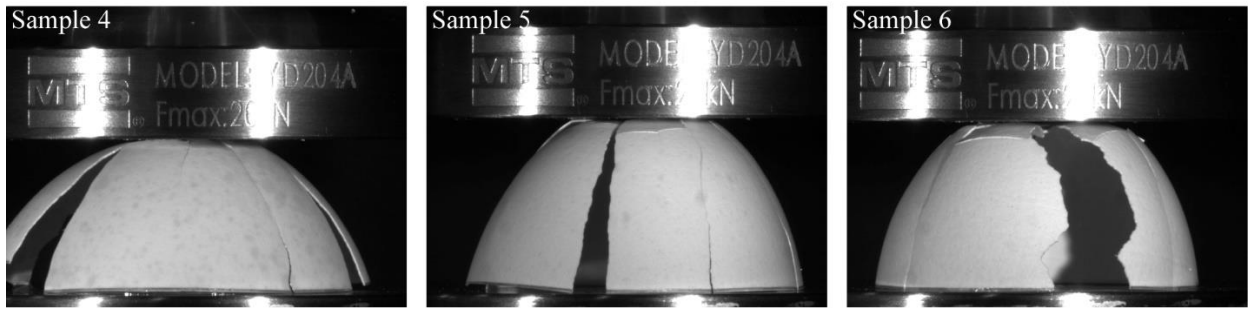

Figure S7. Fracture patterns of eggshell samples with and without membranes.

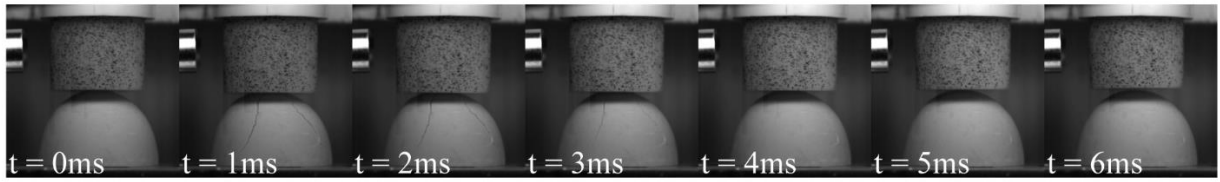

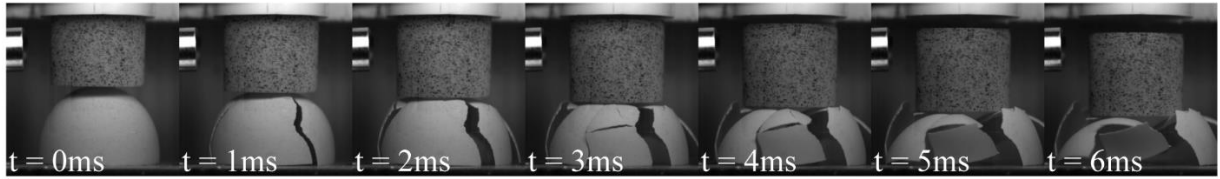

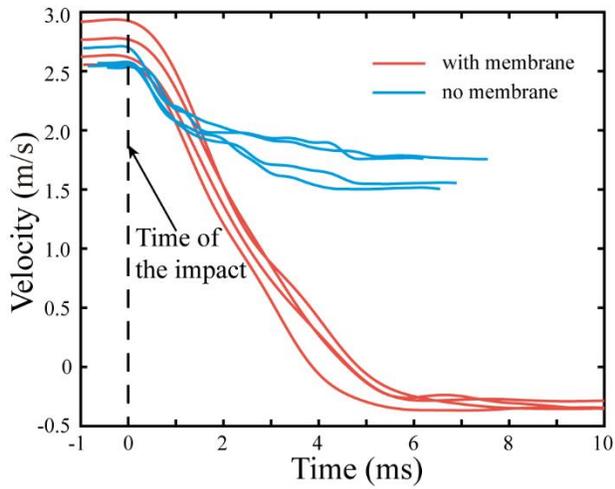
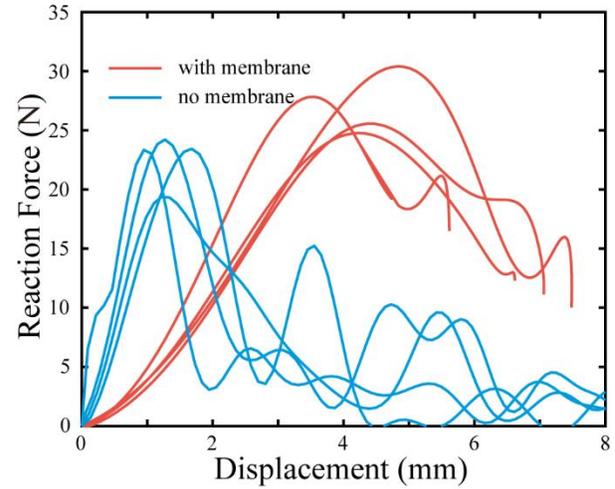

Figure S8. (A-B) Impact tests of the eggshells with and without membranes. (C) The velocity change of the impactor. (D) The reaction force-displacement curves.

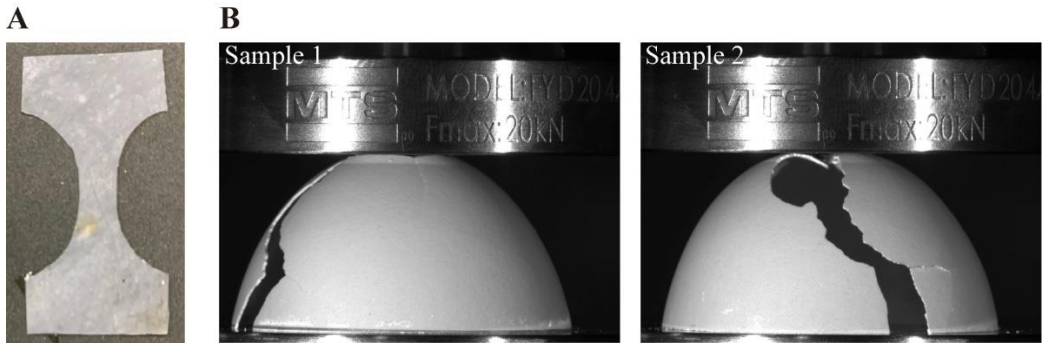

Figure S9. (A) Membrane specimen for tensile test. (B) Fracture patterns of eggshell samples with dry membranes.

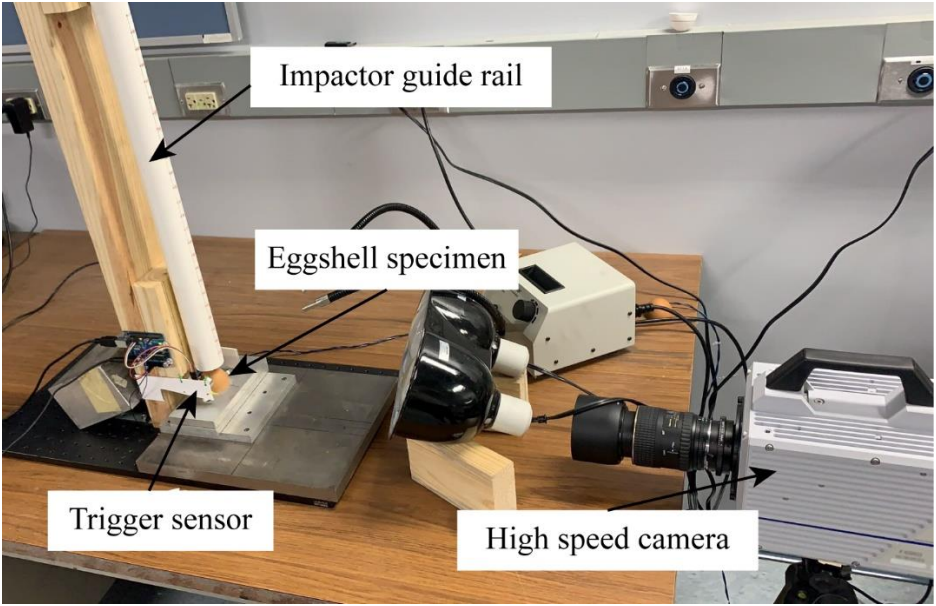

Figure S10. Self-assembled impact test equipment.

Table S1. Geometric parameter of eggshell samples.

| Samples | Geometric parameters | | | |
|---|---|---|---|---|
| | 2a (mm) | 2b (mm) | e (mm) | t (mm) |
| #1 | 54.1 | 44.9 | 2.1 | 0.37 |
| #2 | 54.8 | 44.3 | 1.97 | 0.36 |
| #3 | 56 | 45.2 | 1.99 | 0.38 |
| #4 | 56.8 | 44.4 | 2.03 | 0.34 |
| #5 | 57.3 | 44.9 | 2.05 | 0.37 |
| #6 | 59.1 | 43.2 | 2.07 | 0.39 |
| #7 | 60.2 | 44.3 | 2.13 | 0.37 |
| #8 | 54.7 | 45.2 | 2.18 | 0.35 |
| #9 | 54.8 | 44.6 | 2.24 | 0.36 |
| #10 | 56.5 | 44.7 | 2.25 | 0.33 |
| Mean | 56.4 | 44.6 | 2.10 | 0.36 |
| Standard deviation | 2.0 | 0.6 | 0.10 | 0.02 |

Table S2. Elastic properties of eggshell samples.

| Samples | Elastic properties | | Samples | Elastic properties | |
|---|---|---|---|---|---|
| | $E$ (GPa) | $\nu$* | | $E$ (GPa) | $\nu$* |
| #1 | 33.4 | 0.3 | #11 | 26.6 | 0.3 |
| #2 | 31.1 | 0.3 | #12 | 32.9 | 0.3 |
| #3 | 36.8 | 0.3 | #13 | 29.5 | 0.3 |
| #4 | 25.5 | 0.3 | #14 | 27.0 | 0.3 |
| #5 | 34.7 | 0.3 | #15 | 26.5 | 0.3 |
| #6 | 39.8 | 0.3 | #16 | 25.2 | 0.3 |
| #7 | 25.4 | 0.3 | #17 | 25.3 | 0.3 |
| #8 | 29.5 | 0.3 | #18 | 33.1 | 0.3 |
| #9 | 24.7 | 0.3 | #19 | 31.2 | 0.3 |
| #10 | 24.7 | 0.3 | #20 | 35.4 | 0.3 |
| Mean of $E$ | 30.0 | | | | |
| SD of $E$ | 4.6 | | | | |

*The value of $\nu$ is not from our experiments but from previous studies.

Table S. Tensile strength of eggshell samples.

| Samples | Geometric parameters | | Mechanical properties | |
|---|---|---|---|---|
| | $w$(mm) | $t$ (mm) | Breaking Force (N) | Tensile strength (MPa) |
| #1 | 5.41 | 0.37 | 1.20 | 18.8 |
| #2 | 5.5 | 0.36 | 1.23 | 19.5 |
| #3 | 4.89 | 0.36 | 0.94 | 16.7 |
| #4 | 4.76 | 0.4 | 1.40 | 23.1 |
| #5 | 5.35 | 0.39 | 1.52 | 22.8 |
| #6 | 5.15 | 0.39 | 1.19 | 18.6 |
| Mean | | | | 19.9 |
| Standard deviation | | | | 2.5 |